\documentclass[aps,prd,onecolumn,groupedaddress,superscriptaddress,letterpaper]{revtex4-1}
\usepackage{booktabs}
\usepackage{graphicx}  
\usepackage{color}     
\usepackage{amsmath}   
\usepackage{amsfonts}   
\usepackage{setspace}  
\usepackage{dcolumn}   
\usepackage{enumitem}  
\usepackage{bm}        
\usepackage[usenames,dvipsnames]{xcolor}
\usepackage{geometry}
\usepackage{cancel}

\begin{document}

\title{Thermodynamic potential for quark-gluon plasma with finite quark masses and chemical potential}
\author{Mayukh Raj Gangopadhyay}
\email[]{mayukhraj@gmail.com}
\affiliation{Center for Astrophysics, Department of Physics, University of Notre Dame, Notre Dame, IN 46556}

\author{Grant J. Mathews}
\email[]{gmathews@nd.edu}
\affiliation{Center for Astrophysics, Department of Physics, University of Notre Dame, Notre Dame, IN 46556}

\author{J. Pocahontas Olson}
\email[]{pokie007@gmail.com}
\affiliation{Center for Astrophysics, Department of Physics, University of Notre Dame, Notre Dame, IN 46556}


\date{\today} 

\begin{abstract}
We summarize the derivation of the finite temperature, finite chemical potential thermodynamic potential in the bag-model approximation to quantum chromodynamics (QCD)  that includes a finite $s$-quark mass in the Feynman diagram contributions for both zero-order and two-loop corrections to the quark interaction. The thermodynamic potential for quarks in QCD is a desired ingredient for computations of the equation of state in the early universe, supernovae, neutron stars, and heavy-ion collisions. The 2-loop contributions are normally divergent and become even more difficult in the limit of finite quark masses and finite chemical potential. We introduce various means to interpolate between the low and high chemical potential limits. Although physically well motivated, we show that the infinite series Pad\'e  rational polynomial interpolation scheme introduces spurious poles.  Nevertheless, we show that lower order interpolation schemes such as polynomial interpolation reproduce the Pad\'e result without the presence of spurious poles.  We propose that in this way one can determine the equation of state for the two-loop corrections for arbitrary chemical potential, temperature and quark mass. This provides a new realistic bag-model treatment of the QCD equation of state.  We compute the QCD phase diagram with up to the two-loop corrections.  We show that the two-loop corrections decrease the pressure of the quark-gluon plasma and therefore increase the critical temperature and chemical potential of the phase transition.   We also show, however,  that the correction for finite $s$-quark mass in the two-loop correction serves to decrease  the critical temperature for the quark-hadron phase transition in the early universe. \\
\\
Keywords: thermodynamic potential, quantum chromodynamics, finite quark masses, Feynman diagrams, two-loop corrections,  QCD phase diagram, equation of state for proto-neutron stars.

\end{abstract}

\maketitle

\section{Introduction}
A description of  the equation of state  (EoS) of matter formed in the early universe, heavy-ion nuclear collisions, supernovae, or neutron stars should include the consequences of a possible phase transition between hadronic matter  and quark gluon plasma (QGP). For many astrophysical applications the  description of quark matter has  only been considered in  the zero-order  bag model (e.g. \cite{Alcock87, Fuller88, Gentile93,Sagert09,Fischer10,Fischer11}).   However, the   2-loop corrections at the next order are also significant and should be considered in the EoS  \cite{Kapusta78}. In this approach one  can then construct the QGP EoS from a phase-space integral representation over the scattering amplitudes. Another approach is to model the QGP in the context of a chiral effective field theory (e.g. \cite{Weise12,Buballa16} and refs. therein). Although considerable recent progress has been made in this later approach, here we consider the former bag model approach because it is often employed in astrophysical simulations for its simplicity.

In this approach, it is convenient to compute the EoS for the QGP in terms of the grand thermodynamic potential, $\Omega(T,V,\mu)$ \cite{McLerran86}. Adopting the convention of Landau and Lifshitz \cite{Landau69},  the thermodynamic potential can be  defined in terms of the partition function $Z$ as:
\begin{equation}
\Omega =  -\frac{1}{\beta} \ln{Z}~,
\end{equation}
where $\beta \equiv 1/kT$ (henceforth we adopt units $k = c = \hbar = 1$).   In the Feynman path integral formulation, the partition function is represented as a functional integral of the exponential of an effective action integrated over all fields;
\begin{equation} 
Z = C(\beta) \int [d\phi] \exp{\{i \int_0^{-i\beta} dx_o \int_V d^3 x {\cal L}_{\rm eff}(\phi(x), \partial_\mu \phi(x); {\bf \mu}) \} }~.
\end{equation}
Here, $C(\beta)$ is a normalization and the integration is performed over periodic boson loops and anti-periodic fermion loops.  In quantum chromodynamics the fields $\phi$ are the fermion, gluon, 
and ghost fields \cite{Kapusta78}.

When including up to 2-loop corrections, the grand thermodynamic potential   takes the form:
\begin{equation}
\Omega = \sum_i(\Omega_{q0}^i + \Omega_{q2}^i) +  \Omega_{g0} + \Omega_{g2} + B V~,
\label{eq:GrandPotential}
\end{equation}
where the sum is over quark flavors, (for our purposes $i =$ up, down, and strange).  Here, $q_0$ and $g_0$ denote the $0^\text{th}$-order bag model thermodynamic potentials for quarks and gluons, respectively, while $q_2$ and $g_2$ denote the 2-loop corrections. In the last term $BV$ is the QCD vacuum energy with $B$ the bag constant.  In most calculations sufficient accuracy is obtained by using fixed current-algebra masses. For this work we utilize a strange quark mass of $m_s = 95^ {+9}_{-3}$ MeV, while $m_u = 2.2^ {+0.5}_{-0.4}\sim 0$ MeV and $m_d = 4.7^ {+0.5}_{-0.3} \sim  0$ MeV from the Particle Data Group \cite{PDG18}. 
We also adopt  a bag constant $B^{1/4} = 165 - 240$ MeV as this range should lead \cite{Fuller88} to a critical temperature consistent with the range deduced in the low chemical potential limit from from lattice gauge theory \cite{Kronfeld12}. 
 
With the parameters thus defined, the  quark contribution to the thermodynamic potential is then  given in terms of a sum of the ideal gas contribution plus a two loop correction from phase-space integrals over the Feynman amplitudes~\cite{Kapusta78,McLerran86}:
\begin{widetext}
	\begin{align}
		\label{eqn:idealgas} 
		\Omega_{q0}^{i} =& -2N_cT V\int_0^{\infty} \frac{d^3p}{(2 \pi)^3} 
			\left[\text{ln}\left(1+e^{-\beta\left(E_i-\mu_i\right)}\right) 
				+ \text{ln}\left(1+e^{-\beta\left(E_i+\mu_i\right)}\right)\right]  ~~,  \\
		\label{eqn:twoloop}
		\Omega_{q2}^i =& \alpha_s \pi N_g V\Bigg[
			\frac{1}{3} T^2 \int_0^\infty \frac{d^3p}{(2 \pi)^3}\frac{N_i(p)}{E_i(p)}
				+ \int_0^\infty \frac{d^3p}{(2 \pi)^3} \frac{d^3p'}{(2 \pi)^3}
					\frac{1}{E_i(p)E_i(p')}\left[N_i(p)N_i(p')+2\right] 	\nonumber \\
			  &\qquad \qquad \quad \times \left[
			  \frac{N_i^+(p)N_i^+(p')+N_i^-(p)N_i^-(p')}{\left(E_i(p)-E_i(p')\right)^2-\left({\bf p-p'}\right)^2} 
			  + \frac{N_i^+(p)N_i^-(p')+N_i^-(p)N_i^+(p')}{\left(E_i(p)+E_i(p')\right)^2-\left({\bf p-p'}
			  \right)^2}\right]\Bigg] ~~,
\end{align}
\end{widetext}
where the sum on $i$ is over quark flavors, $N_c = 3 $ is the number of colors,  $N_g= 8$ is the number of gluons, and $E_i(p) = \sqrt{m + p^2}$ is the relativistic energy.   The $N_i^{\pm}$ denote the quark and anti-quark Fermi-Dirac distributions:
\begin{equation}
N_i^{\pm}(p) = \frac{1}{e^{\beta\left(E_i(p)\mp\mu_i\right)} + 1}   ~.
\label{fdirac:eq}
\end{equation}

The one- and two-loop gluon and ghost contributions to the thermodynamic potentials can be evaluated in a similar fashion to that of the quarks.
\begin{align}
\Omega_{g0} =& 2 N_g T V\int_0^\infty\frac{d^3p}{(2 \pi)^3}\text{ln}\left(1-e^{-\beta\lvert p \rvert}\right) \nonumber \\
                         =& -\frac{\pi^2}{45}N_gT^4   ~. 
\end{align}
\begin{equation}
\Omega_{g2} = \frac{\pi}{36}\alpha_sN_c N_gT^4.
\end{equation}

For massless quarks, \mbox{Eqs.~(\ref{eqn:idealgas}-\ref{eqn:twoloop})} are easily evaluated \cite{McLerran86} to give
\begin{align}
\Omega_{q0}^i =& -\frac{N_c V}{6} \left(\frac{7\pi^2}{30}T^4 + \mu_i^2T^2 +\frac{\mu_i^4}{2\pi^2}\right) ~~,\\
\Omega_{q2}^i =& \frac{N_g \alpha_s V}{8\pi}  \left(\frac{5\pi^2}{18}T^4 + \mu_i^2T^2 +\frac{\mu_i^4}{2\pi^2}\right)  ~~,
\label{q20}
\end{align}
where $\alpha_s$ is the strong coupling constant.  Here, one can immediately see from the pre-factors that the 2-loop contribution is comparable to the 0-order contribution.  It is only suppressed by a factor of 
$\sim 2 \alpha_s/\pi$.  For the application here we adopt the low-energy value of $\alpha_s = 0.33$ \cite{PDG18}.

For the massive strange quark the ideal gas contribution [Eq.~(\ref{eqn:idealgas})] can be easily integrated as described below. However, the two-loop correction [Eq.~(\ref{eqn:twoloop})] cannot be integrated numerically, due to inherent divergences. It is,  therefore, common to ignore the 2-loop correction or to approximate the two loop strange quark contribution with the zero mass limit \cite{McLerran86}. 

This may, however,  over estimate the strange-quark contribution as it ignores the Boltzmann factor suppression of the thermodynamic potential for quarks with finite mass.  When the quark mass is relatively small compared to its chemical potential, this may be a reasonable approximation.  However, this is not necessarily the case as one approaches the QCD phase transition at moderate values of the chemical potential.

 In this paper, therefore, we attempt to extend this zero-mass quark approximation to finite masses and chemical potential followed by an analytic extension to the massless limit.  
 We propose that this provides  a more realistic input to the equation of state than the zero-quark mass limit for the 2-loop diagrams that is usually employed.  As an illustration we compute the bag-model QCD phase diagram for a finite-mass $s$-quark and contrast this with the massless limit.
 
 \section{Feynman Diagrams}
 
\begin{figure}[t]
    
  \centering
 \includegraphics[width=0.5\hsize]{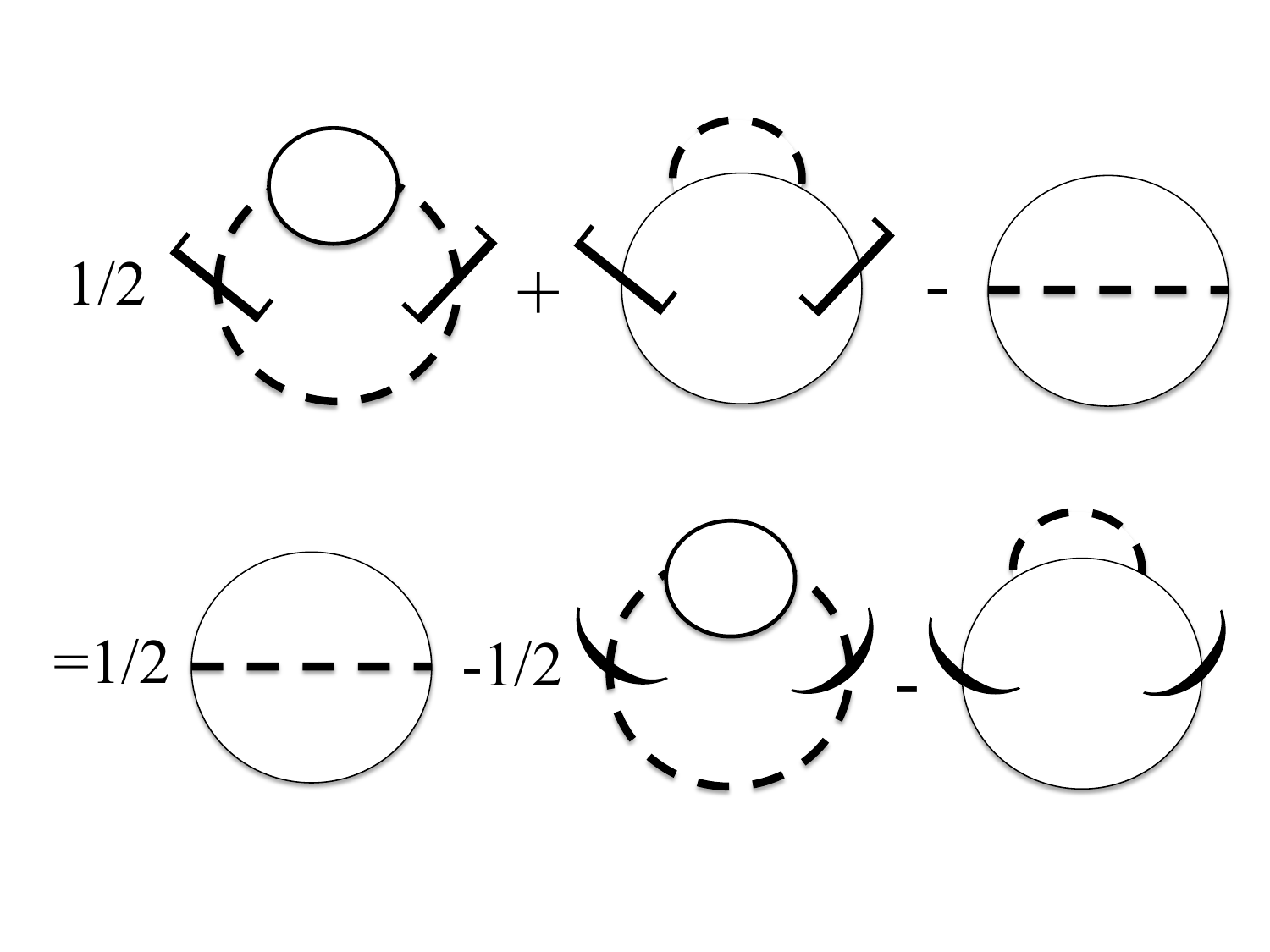}
  \caption{Second order fermion 2-loop contributions to the thermodynamic potential.}
  \label{fig:1}
    
   \end{figure}

 The Feynman diagrams for the fermions that make the 2-loop contribution to the thermodynamic potential are shown in Figure \ref{fig:1}.
However, it is a well known problem that for finite mass all of the Feynman diagrams diverge as $\Lambda^4$  as  the ultraviolet cutoff increases.   
Here, however, one can slightly circumvent this problem by introducing periodic regularization as in \cite{Kapusta78}. That is, the quark masses are individually treated with periodic cut-offs.

\begin{figure}[t]
    
  \centering
 \includegraphics[width=0.6\hsize]{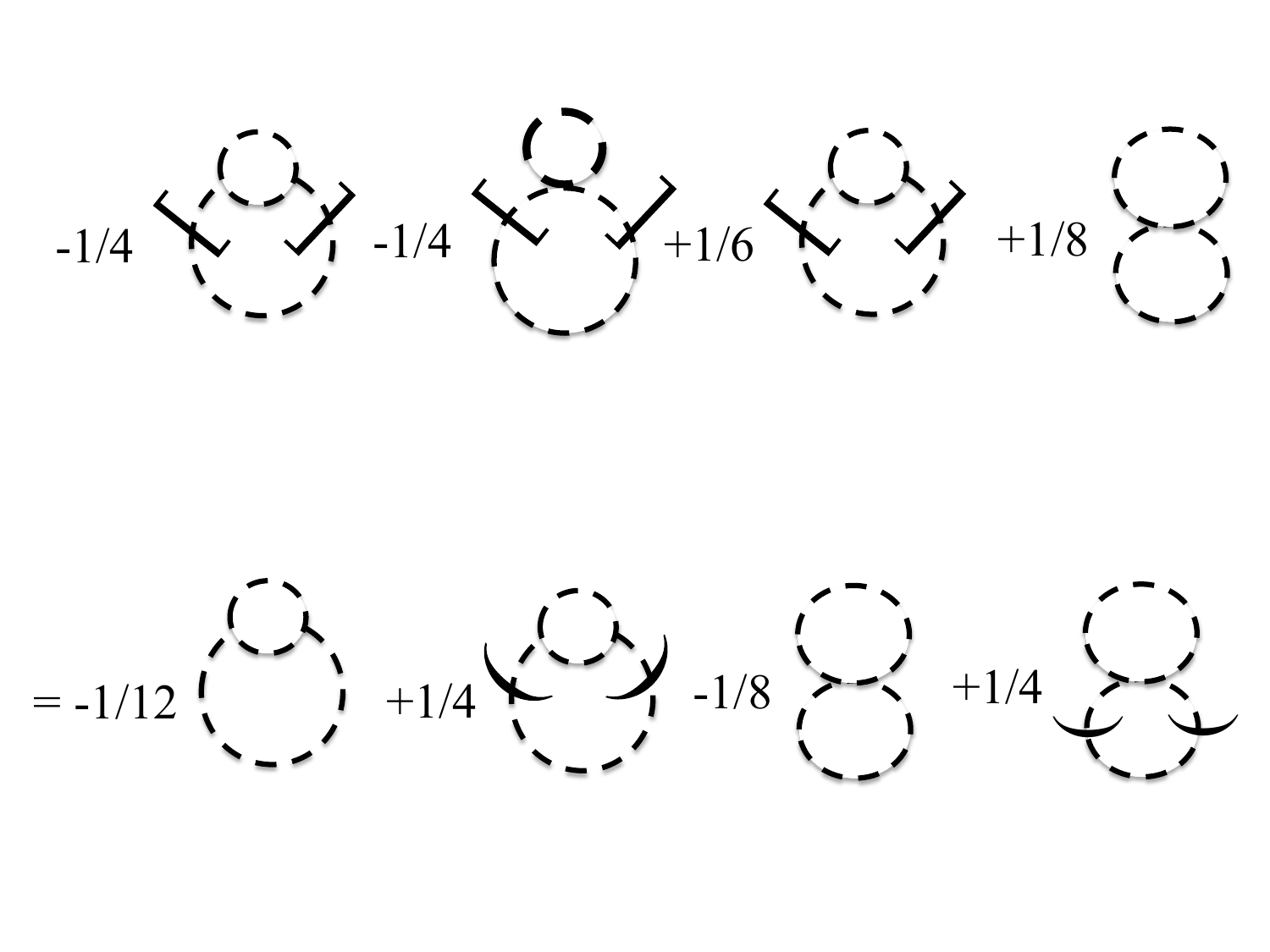}
  \caption{Second order gluon ghost 2-loop contributions to the gluon thermodynamic potential.}
  \label{fig:2}
   \end{figure}

The 2-loop gluon Feynman diagrams that contribute to the potential due to  the ghost loops are shown in Figure \ref{fig:2}.

\section{Method of Calculation}
It is generally expected that for sufficiently high densities and/or temperature, a transition from hadronic matter to quark-gluon plasma (QGP) can occur~\cite{McLerran86}. Progress in lattice gauge theory (LGT) has shed light on the transition to a  QGP in the low baryon chemical potential, high-temperature limit~\cite{Kronfeld12}. It is now believed that at high temperature and low density a deconfinement and chiral symmetry restoration occur simultaneously at the crossover boundary. In particular, at low density and high temperature, it has been found that the order parameters for deconfinement and chiral symmetry restoration changes abruptly for critical temperatures of \mbox{$T_c = 145 - 170$ MeV}~\cite{Kronfeld12, Borsanyi12, Bazavov12}.  
However, neither order parameter exhibits the characteristic change expected from a 1st order phase transition. Indeed, an analysis of many  thermodynamic observables confirms that the transition from a hadron phase to a high temperature QGP is a smooth crossover~\cite{Aoki06,Bazavov09}. However, it is expected that for higher chemical potential $\mu ^>_\sim 300$ MeV, a critical point appears at which first order chiral transition can occur \cite{Kronfeld12}. We note that the bag-model described here does not distinguish between the chiral and deconfinement transition.  Hence, we do not describe the critical point deduced from LGT, but take that as a given.   Moreover, the order of the transition requires a determination of the surface tension for nucleated bubbles of QGP \cite{Fuller88, Mariani17}.  However, there is considerable uncertainty in the surface tension. Hence, we assume an abrupt transition between hadronic and quark matter as in Ref.~\cite{Mariani17}.

In spite of these limitations, it is nevertheless worthwhile for astrophysical applications to evaluate  the impacts on the QCD phase diagram  from both the zero-order bag model plus  the massless two-loop contributions, along with effects of the finite $s$-quark mass on the 2-loop contribution.  That is our goal in this paper.

\subsection{Evaluation of Feynman Diagrams}
The evaluation of the fermion Feynman diagrams shown in Figure \ref{fig:1} can be reduced to the following  the contour integral~\cite{Kapusta78},
\begin{widetext}
\begin{align}
		\label{eqn:iinteg} 
\Omega_{q2} =  \frac{2 \pi \alpha_s N_g V}{2} \int \frac{d^3p}{(2\pi)^3} \frac{d^3q}{(2\pi)^3} \frac{d^3k}{(2\pi)^3} (2 \pi)^3 \delta(p - q - k) \nonumber\\
\times T^3 \sum_{n_p n_q n_k} \beta \delta_{n_p, n_q + n_p}\frac{Tr\biggl[ \gamma^{m} (\cancel{p}+m) \gamma_m(\cancel{q} + m)\biggr]}
{k^2(p^2 - m^2)(q^2 - m^2)}  \\
 = -8T^3  \sum_{n_p n_q n_k} \beta \delta_{n_p, n_q + n_p}\frac{2m^2 - p\cdot q}
{k^2(p^2 - m^2)(q^2 - m^2)} ~~.
\end{align}
\end{widetext}
One can  then evaluate the $\beta \delta_{n_p, n_q + n_p}$ factor using periodic conditions to regularize the function:
\begin{eqnarray}
\beta \delta_{n_p, n_q + n_p} &= &\int_0^\beta \exp{[\textbf{i}\Theta(p^0 - q^0 - K^0)]} \nonumber \\
 &=& \frac{\exp{[\textbf{i}\beta(p^0 - q^0 - K^0)]} - 1}{p^0 - \gamma^0 - K^0} ~~,
 \end{eqnarray}
where, the periodicity obeys,
\begin{equation}
\beta^0 = (2 n_p + 1) \prod_i T_i + \mu~~,
\end{equation}
\begin{equation}
q^0 = (2 n_q + 1) \prod_i T_i + \mu ~~,
\end{equation}
\begin{equation}
K^0 = 2 n_k \prod_i T_i ~~,
\end{equation}
where $T_i$ is the function to regularize.  Finally, using the on-shell condition,
\begin{eqnarray}
I(p^0, q^0, k^0) &=& \frac{2 m^2 - p \cdot q}{p^0 - q^0 - k^0} \nonumber \\
&\times& \biggl( \exp{[\beta(k^0 + q^0 - \mu)]} - \exp{[\beta(p^0 -  \mu)]}  \biggr)~~,
\end{eqnarray}
and performing the contour integral one has the desired result,
\begin{eqnarray}
\Omega_{q2}  &=&  \frac{1}{6} 2 \pi \alpha_s N_g V T^2  \int \frac{d^3p}{(2\pi)^3}  \frac{N_F(p)}{E_p}
 \frac{1}{2} 2 \pi \alpha_s N_g V \int \frac{d^3p}{(2\pi)^3} \frac{d^3q}{(2\pi)^3} \nonumber \\
&\times&\frac{1}{E_p E_q} \biggl[ \biggl( 1 + \frac{2 m^2}{(E_p - E_q)^2 - (p - q)^2} \bigr) \bigl[ N_F^- (p) N_F^-(q)  + N_F^+ (p) N_F^+(q) \bigr] \nonumber \\
& +& \bigl[ N_F^+ (p) N_F^-(q)  + N_F^- (p) N_F^+(q) \bigr]\biggr]  ~~,\nonumber \\
\label{eqn:iinteg2} 
\end{eqnarray}
where, $N_F^+ $ and $ N_F^-$ are the fermion distribution functions as defined in Eq.~(\ref{fdirac:eq}).

In the zero-temperature limit then,
\begin{widetext}
\begin{align}
		\label{eqn:iintegT0} 
\Omega_{q2}(T=0) = \frac{2 \pi \alpha_s N_g}{(2 \pi)^4} \biggl( \frac{3}{2} \biggl[ \mu p_f - m^2 \ln{\biggl( \frac{\mu + p_f}{m}\biggr)} \biggr]^2 - p_f^4\biggr)
  \int \frac{d^3p}{(2\pi)^3}  \frac{N_F(p)}{E_p} ~,
\end{align}
\end{widetext}
 where, $N_F$ is also defined in Eq.~(\ref{fdirac:eq})  for ghosts.
In the finite temperature $m \rightarrow 0$ limit we then have:
\begin{equation}
\Omega_{q2}(m=0) =  \frac{2 \pi \alpha_s N_g T^4 }{288} \biggl( 5 + \frac{18}{\pi^2 }\frac{\mu^2 }{T^2} + \frac{9}{\pi^4} \frac{\mu^4}{T^4} \biggr)~~,
\label{zerom}
\end{equation}
which trivially reduces to Eq.~(\ref{q20}).

Finally, for moderate temperature and  $ \mu \le m $ we have the result of interest for this paper.
\begin{equation}
\Omega_{q2}(T,m,\mu) =  \frac{\alpha_s N_g V}{8\pi^2} m^2 T^2 \exp{ [2 (\mu - m)/T]}~~.
\label{finitem}
\end{equation}
This expression obviously diverges exponentially for $\mu > m$ and diverges with temperature as $T^2$. Nevertheless, it is useful for low to moderate chemical potential and temperatures up to and beyond $\sim m_s $.
Moreover, in the next section we describe  an analytic interpolation between the massive and massless regimes that we propose is a more accurate representation of the true EoS than to assume massless quarks for the two-loop correction.

\section{Results}
\begin{figure}[t]
  \centering
 \includegraphics[width=0.9\textwidth]{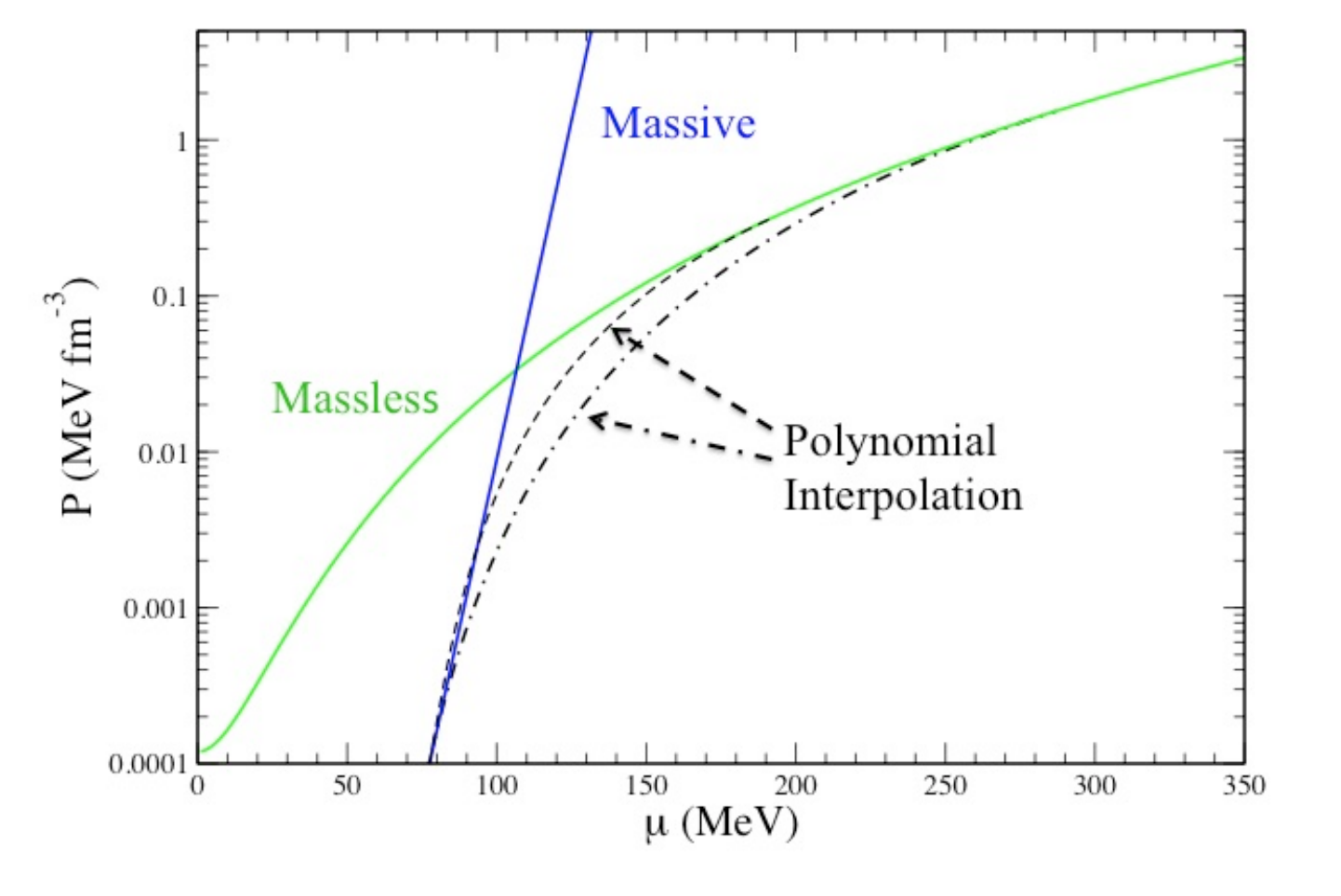}
\caption{(Color online) Two loop contribution to the $s$-quark pressure as a function of chemical potential.  Green line is based  upon the massless limit  of Eq.~(\ref{zerom}). The blue line shows the results of Eq.~(\ref{finitem}) for $m_s = 95$ MeV.  Black lines show an interpolation between the regimes  for a 5th order polynomial between $\mu = 65$ to $285$ MeV (dot-dashed  line) and $\mu = 75$ to $190$ MeV (dashed line).}
  \label{fig:3}
 \end{figure}

In Figure \ref{fig:3} we compare calculations of the magnitude of the the $s$-quark 2-loop contribution to the pressure as a function of chemical potential,  where
\begin{equation}
P = -\biggl( \frac{\partial \Omega}{\partial V} \biggr)_{T, \mu} ~~.
\end{equation}
Here, we take a realistic $s$-quark mass of $m_s = 95$ MeV and select a typical core-collapse supernova  temperature of $T=10$ MeV. Lines are drawn for calculations  in the $m_s = 0$  limit (green line) and in the  finite-mass  periodic-regularization of Eq.~(\ref{finitem}) (blue line). For finite  $s$-quark mass, there is an obvious divergence at large $\mu/T$ due to the $\exp{[ 2 \mu/T]}$ term in Eq.~(\ref{finitem}).  This limits the range of validity of  this periodic regularization approach.  
Moreover, as can be seen in the example given in Figure \ref{fig:3}, for low chemical potential, $\mu < m_s$, the massless approximation over-estimates the pressure contribution due to the $s$-quark by more than an order of magnitude, and continues to overestimate the pressure by at least a factor of two up to $\mu \sim 1.5 m_s$.

Thus, one can only apply Eq.~(\ref{finitem}) in the low chemical-potential limit, i.e. $\mu  \le m_s$.  On the other hand, in the large chemical potential limit, $\mu ^>_\sim  2 m_s$ one expects the thermodynamic potential  to approach the massless limit of Eq.~(\ref{zerom}). So, we are faced with the dilemma that the function is well posed for $\mu < m_s$ and  $\mu ^>_\sim 2 m_s$, but is not well defined between.
Hence, we have considered various interpolation schemes.
 
\subsection{Interpolation Schemes}
To interpolate between the small and large chemical potential limits of the EoS a Pad\'e rational polynomial interpolation might seem physically well motivated.  Indeed, there is a vast literature dealing with the Pad\'e approximants and  there exist many examples in physics of  quantities that can be deduced by Pad\'e approximants
 \cite{Baker81,Press07,Vatsya09}.  For example, the $D$-Pad\'e method provides a means to extrapolate from the weak coupling to strong coupling limits of Hamiltonian Lattice QCD in the $t$-expansion \cite{Horn85, Mathews87}.  Our desire here is to reconstruct the infinite series that connects between regimes of low  to high values of $ \mu/T$.  Hence, we seek guidance in the construction of that infinite series by Pad\'e approximants.  
  
 The Pad\'e  approximation consists of a rational polynomial to produce an infinite series that is often a better approximation to a function than truncating its Taylor series, and may still work even in cases where the Taylor series does not converge. In the application of interest here, the Pad\'e series involves well defined approximate functions in the low and high limits of the variable $x = \mu/T$. 
 
 There is a well established technique for numerical Pad\'e interpolation between the ranges of validity  \cite{Press07}. This is based upon the algorithm of Burlisch-Stoer \cite{Stoer80} that generates rational functions through  a recursion relation based upon Neville's algorithm that makes use of data that exist at the two limits of the interpolation \cite{Press07}.
 
\begin{figure}[t]
  \centering
 \includegraphics[width=0.9\textwidth]{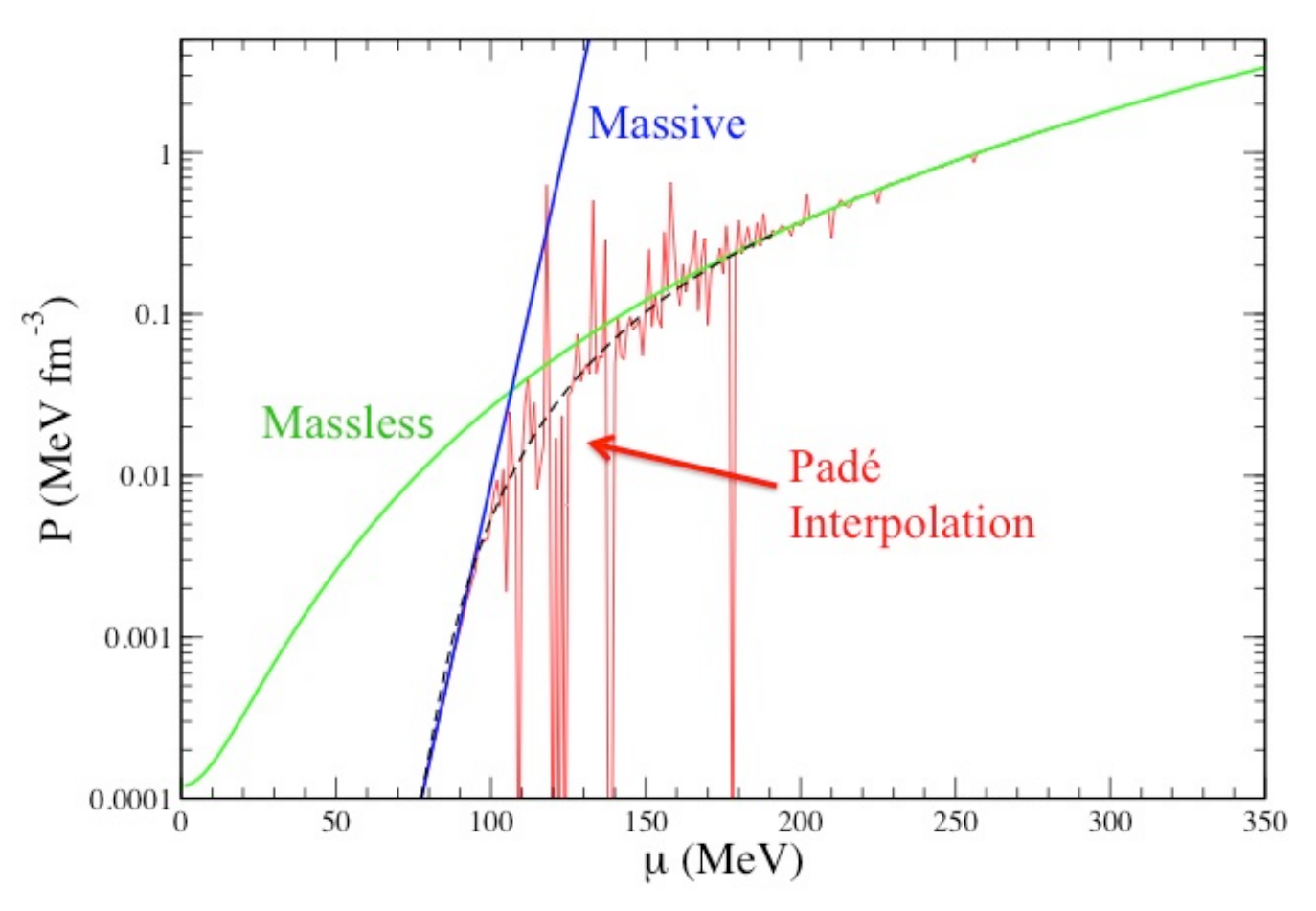}
  \caption{Same as Figure 3, but in this case, the redline  shows a Pad\'e  interpolation between the regimes from  $\mu = 75$ MeV to  $\mu =190$ MeV.  For comparison the dashed line shows  a 5th order polynomial interpolation in the same interval.  Clearly, spurious poles affect the reliability of Pad\'e interpolation, although on average it follows the polynomial interpolation.}
  \label{fig:4} 
   \end{figure}

Figure \ref{fig:4} illustrates the numerical Pad\'e interpolation \cite{Press07} based upon this algorithm  in the range from from $\mu = 75$ MeV to  $\mu =190$ MeV.
As can be seen by the red line in Figure \ref{fig:4}, the Pad\'e interpolation unfortunately   leads to a number of spurious poles in the Pad\'e function.  Indeed,  spurious poles in the Pad\'e function are a common feature of rational polynomial interpolation  \cite{Stahl98}.  Although methods have been discussed to minimize them  \cite{Baker81,Vatsya09}, we have found these to be difficult to implement and not particularly useful.    
 In the present application these spurious poles can be traced to the fact that the application here differs from a simple truncated Taylor series. We have explicit functions in both the low chemical potential $m >> \mu$  [Eq.~(\ref{finitem})] and large $m << \mu$ limit [Eq.~(\ref{zerom})].  Although the lower limit  limit can be represented by a Taylor-Maclaurin series around  $x = (\mu/T) =0$, the upper limit is a simple quartic function.  Achieving this limit requires a cancellation of many terms in the infinite Pad\'e series in the limit of large $\mu/T$.  
   
 Hence, it is difficult to impose a rational polynomial that asymptotes to Eq.~(\ref{zerom}) without introducing spurious poles in the recursion relation.  Nevertheless, the correct interpolation is evident in the trend represented in the Pad\'e approximant.  The next obvious choice is that of a simple polynomial interpolation scheme.  However, because of the rather steep transition between the behavior at low and high chemical potential a rather high order polynomial is needed.
        
        The dashed line on Figures \ref{fig:3} and  \ref{fig:4} shows a 5th-order polynomial  interpolation between the same limits as the Pad\'e series.  One can see that a polynomial interpolation based upon the same Neville  recursion relation \cite{Press07} follows the trend of the Pad\'e interpolation without the introduction of spurious poles.  Hence,  a simple 5th order polynomial interpolation between the regimes obtains  a result that is equivalent to the Pad\'e interpolation but avoids  the spurious poles.

\begin{figure}
  \centering
 \includegraphics[width=0.9\textwidth]{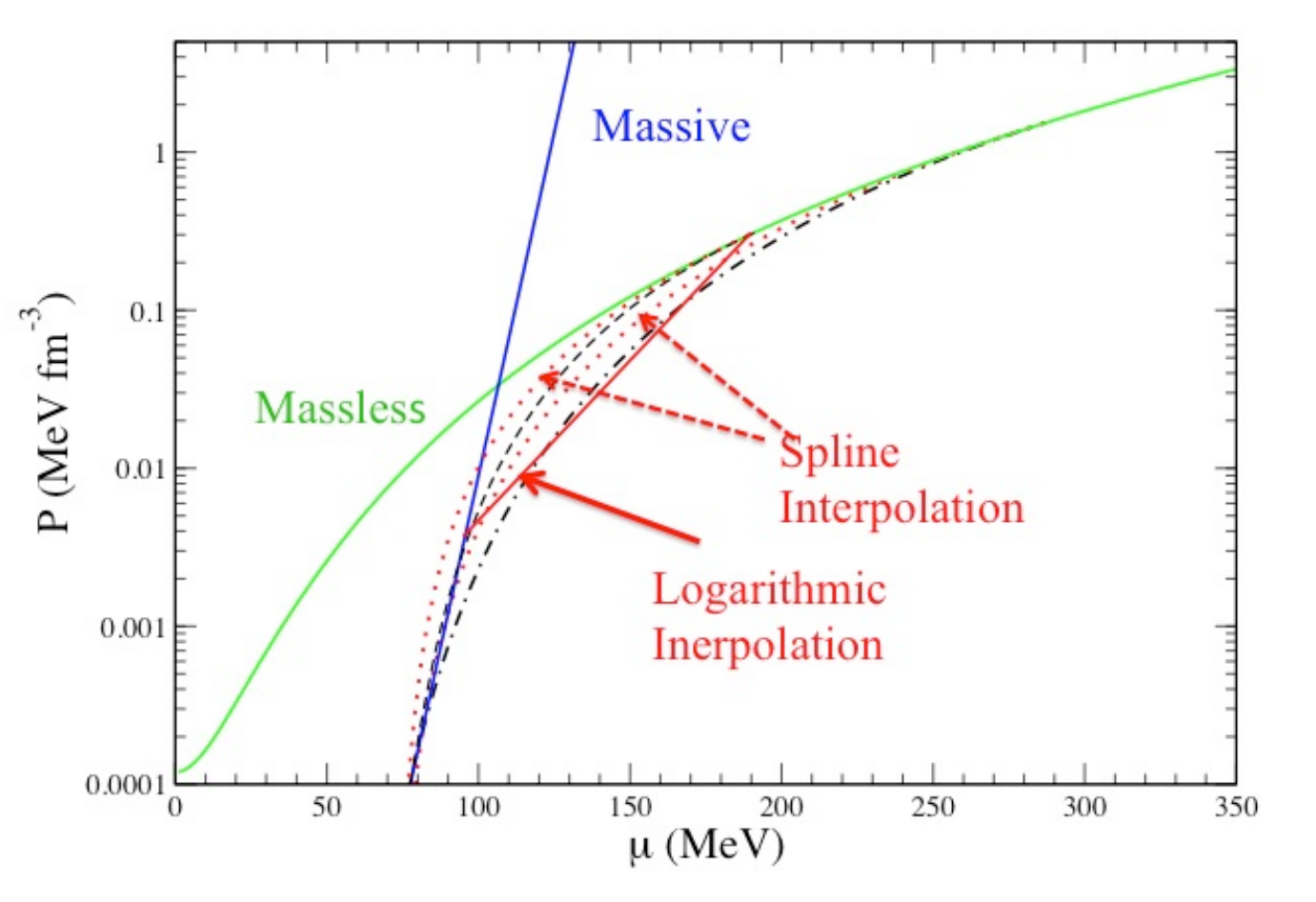}
  \caption{Same as Figure 3, but in this case, the red dotted lines line  show a cubic spline   interpolation between the regimes from  $65 < \mu < 285$ MeV (upper curve) 
  and $75 < \mu < 190$ MeV (lower curve).  For comparison the dashed and dot-dashed lines  show  the  5th order polynomial interpolations as discussed in Figure \ref{fig:3}. The spline interpolation is not a good match to  the polynomial or Pad\'e   interpolation schemes. The red solid line shows a simple logarithmic interpolation between $95 \le \mu \le 190$ MeV as discussed in the text.}
  \label{fig:5} 
   \end{figure}

  As one more possibility, we have considered the application of a simple cubic spline  between the small $\mu$ and large $\mu$ regimes \cite{Press07}.  This interpolation is shown on Figure \ref{fig:5} and compared with the polynomial interpolations of Figure \ref{fig:3}.  Here we see, that although a spline interpolation is easier to implement, it does not make use of the entire data set as in the Polynomial recursion relation.  Hence, it does well represent the Pad\'e or  polynomial interpolations.  For this reason, we do not recommend a spline interpolation.

Although a 5th order polynomial interpolation is a possible  means to generate the 2-loop contribution,
  we have found that this choice is difficult to implement in practical calculations.  Indeed, when applied over many different temperatures a 5th order polynomial can lead to spurious fluctuations as serious as the  poles in the numerical Pad\'e method.   Moreover, as can be seen in Figures \ref{fig:4} and \ref{fig:5}, there is inherent uncertainty of about $\pm 30\%$ due to the ambiguity in the choice of when to begin and end the interpolation between the massive and massless limit.   For this reason (and for the fastest practical numerical applications) we suggest a simple logarithmic interpolation between $\mu = m_s$ and $\mu = 2 m_s$.  This gives a comparable uncertainty to that of the best polynomial fits and avoids completely the dangers of spurious poles and wiggles in the interpolation scheme.   This is illustrated by a straight red line on Figure \ref{fig:5}.

\subsection{Impact of present work on the QCD phase diagram}
 There are three aspects of the present work of physical relevance:  1)  On the one hand, we have made a study of the importance of two-loop corrections to the QGP EoS. These are shown to be significant even in the massless limit; 2) On the other hand, we have discussed  the fact that  including the mass of the $s$-quark in the two-loop correction is  divergent, but possible for low to moderate chemical potential  by introducing periodic regularization of the  relevant Feynman diagrams;  and 3)   We have shown, however, that there is a natural way to extrapolate from the low to high chemical potential regime that allows for the inclusion of  the finite $s$-quark mass.   Here we discuss these  aspects and how they affect the QCD phase diagram, i.e. a plot of the critical temperature $T_c$ vs. the baryon chemical potential for the transition between the hadron phase and quark-gluon plasma.

 The construction of the phase diagram requires a model for both the QGP phase described here and the confined hadron phase.  
The equation of state for the hadron phase at very high chemical potential and density relevant to cold neutron stars, however,  requires a detailed treatment of the nuclear-matter equation of state at high density.  We will address this in a future work  following the approach outlined in  \cite{Olson19}. However, our main interest here is the effects of the $s$-quark mass at low to moderate chemical potential. For this purpose one can treat the hadron phase as a non-interacting gas of baryons and mesons that obey the usual Fermi-Dirac or Bose-Einstein statistics.  For this purpose  we can expand the relevant Fermi integrals in terms of modified Bessel functions so that the thermodynamic potential for baryons with $\mu ^<_\sim  m_b \sim 900$ MeV  can be written \cite{Alcock87,Fuller88}:
 \begin{equation}
 \Omega_b = \sum_{i = {\rm baryons}} -\frac{g_i T^4 V}{\pi^2} \sum_{n=1}^{\infty}\frac{(-1)^{n+1}}{n^4} \cosh{(n \mu/T)} \bar K_2( n m_i/T) ~~,
 \label{omegab}
 \end{equation}
 where $\bar K_2(x)$ is related to the modified Bessel function of second order [$\bar K_2 = (x^2/2) K_2$], and the sum is over all baryon resonances listed in the particle data book \cite{PDG18},  with $g_i$ the usual spin factor.
 Similarly, for mesons one can write:
 \begin{equation}
 \Omega_m = \sum_{i = {\rm mesons}} -\frac{g_i T^4 V}{\pi^2} \sum_{n=1}^{\infty}\frac{1}{n^4}  \bar K_2( n m_i/T) ~~.
 \label{omegab}
 \end{equation}
 From the above thermodynamic potentials one can deduce the critical temperature for the sharp transition between the QCD and hadronic phases via a Maxwell construction \cite{Fuller88}.
 
\begin{figure}
  \centering
 \includegraphics[width=0.9\textwidth]{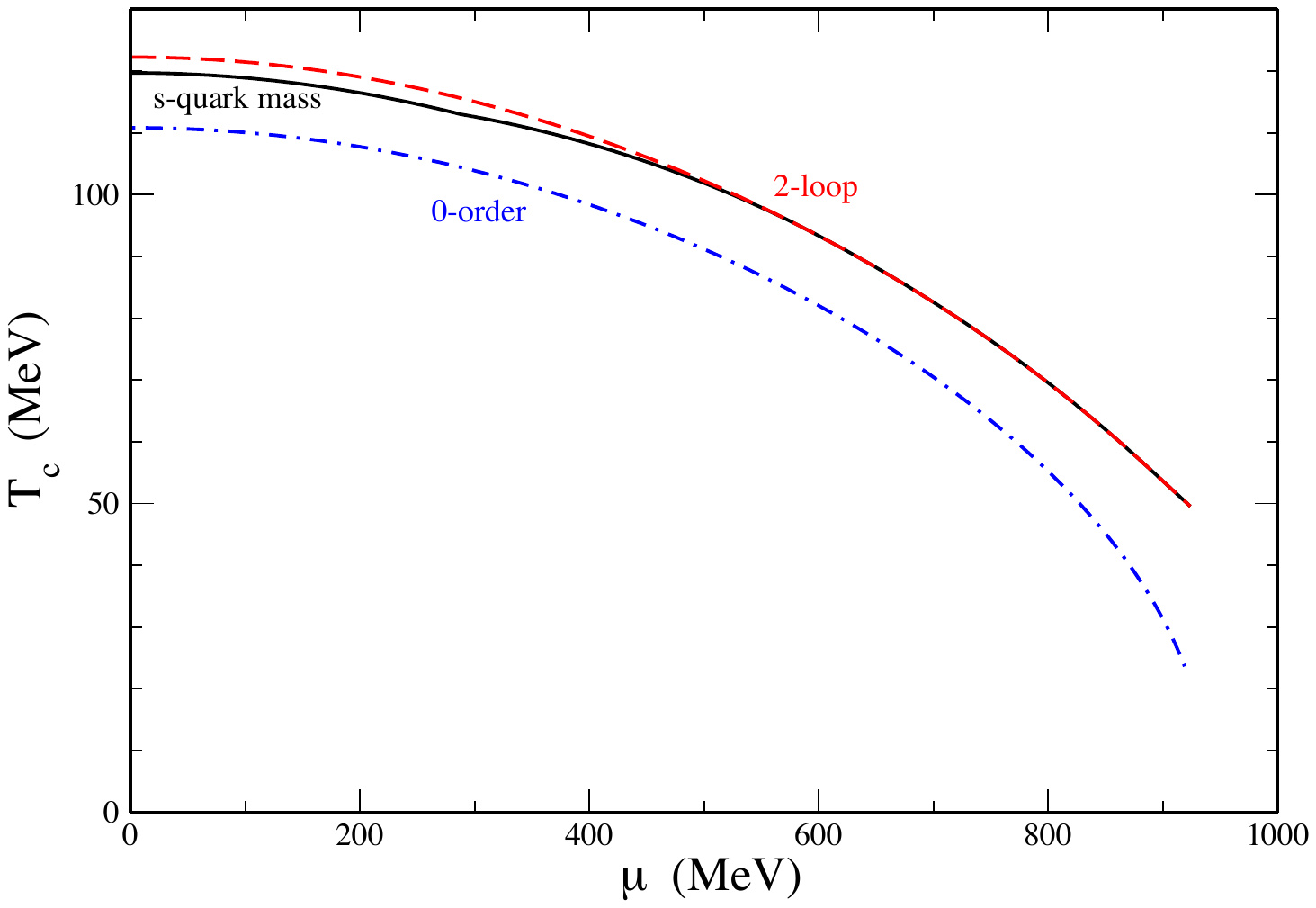}
  \caption{QCD phase diagram relevant to the present work.  Plot shows the critical temperature $T_c$ vs. baryon chemical potential for 3 cases based upon a bag-model QCD vacuum energy of $B^{1/4} = 165$ MeV.  1) Dot-dashed (blue) line shows the usually applied lowest order bag-model prediction for 3 massless quarks; 2) Dashed (red) line shows the importance  of adding the 2-loop corrections with 3 massless quarks;  3) Solid (black) line shows the effect of including the finite mass of the $s$-quark (95 MeV) for the two-loop contribution via the procedure outlined in this paper.   }
  \label{fig:6} 
   \end{figure}

  Figure \ref{fig:6}  shows an example of the impact of the formulation presented here on the QCD phase diagram.  For this illustration, we chose a bag constant ($B^{1/4} = 165$ MeV).  This leads to a high-temperature phase transition   on the low end of the range (145 MeV $\le T_c \le$ 170 MeV) suggested by lattice QCD.  We choose this value because the phase transition for lower values of B are more likely to be manifest in astrophysical environments.   We note, that since we do not include hadronic interactions in the nuclear equation of state, and the modified bessel function expansion is only valid up to $\mu ^<_\sim 900$ MeV, we do not compute the phase diagram beyond 900 MeV. This is adequate for our purpose.
  
  The blue dot-dashed line in Figure \ref{fig:6} shows the effect of the usually adopted  assumption of only the zero-order MIT Bag model   \cite{Gentile93,Sagert09,Fischer10,Fischer11}.   The importance of adding the two-loop contribution in the massless limit is shown by the red dashed line.  Here it is evident that the 2-loop contribution raises the critical temperature and/or chemical potential of the QCD phase transition.  The reason for this is easy to understand.  The 2-loop contribution to the thermodynamic potential enters with the opposite sign. The effect of the 2-loop diagrams is, therefore,  to decrease the pressure in the QGP phase relative to that of the 0-order contribution.  Since the pressure is lower one must go to higher temperature or chemical potential before the pressure of the QGP phase exceeds that of the hadron phase.  Hence, a higher critical temperature results.  Indeed, for this value of the bag constant, the 2-loop correction is required to obtain a critical temperature $T_c \ge 145$ MeV as suggested by LGT.

\subsection{$s$-quark mass and the high-temperature phase transition}

The black line on Figure \ref{fig:6}  shows  the impact of incorporating the finite $s$-quark mass (95 MeV) into the 2-loop contribution via the procedure of periodic regularization and interpolation as discussed in the preceding sections.     As expected, this correction  mainly affects the regime of low chemical potential $(\mu ^<_\sim  300$ MeV) for which the periodic renormalization and extension described above should be valid.  Here one can see that adding a finite $s$-quark mass slightly decreases critical temperature.  This is easy to understand as the effect of the finite mass is to decrease the 2-loop contribution from the $s$-quark  until the chemical potential exceeds the $s$-quark mass.   Thus, the diminished magnitude of the 2-loop contribution means that the pressure of the QGP phase can exceed that of the hadrons for a slightly lower critical temperature.

The portion of the QCD phase diagram in the regime of low chemical potential affected by the $s$-quark mass corresponds to the high temperatures of the early universe or perhaps in heavy ion collisions.   
 There have been a number of studies of the QCD phase diagram and its effect on neutron stars in the context of chiral perturbation theory  (e.g.  \cite{Weise12,Buballa16}).  It is expected that the finite-density QCD phase transition is a first-order chiral transition that occurs for baryon chemical potential from $\mu \sim 300-900 $ MeV.  During a core collapse supernova the central core approaches and exceeds nuclear matter density with $\mu \sim 900$ MeV, while the core temperature can be in excess of 50 MeV.  As the core experiences a first order phase transition the baryon pressure support diminishes (although some electron pressure support remains). The initial approach to nuclear matter density induces a shock as the EoS stiffens  \cite{Wilson03}. However, the subsequent onset of a first-order transition leads to a second shock as the core re-collapses through the phase transition to form pure quark-gluon plasma  \cite{Gentile93,Sagert09,Fischer10,Fischer11}.

 During the initial collapse, the temperature is high enough that the massless limit is appropriate for the $u$ and $d$ quarks.  The production of $s$ quarks will be delayed by the time scale for weak decays, and will be further suppressed by adopting the EoS discussed here due to the Boltzmann factor $\exp{(-m/T)}$.  This will cause the initial evolution through the first and second shock to proceed as though only two massless quark flavors are present.  It is only during the subsequent cooling phase where the higher chemical potentials and production of massive $s$-quarks will be manifest.  
 
 The presence of negatively charged  $s$ quarks will neutralize the matter and diminish the isospin asymmetry.  A large abundance of $s$ quarks  leads to color superconductivity in a color-flavor locked (CFL) phase in which $u$, $d$, and $s$ quarks are paired in a symmetric and electrically neutral way.  This will further decrease the pressure support of the core due to the decreased quark degeneracy pressure as $u,d$ quarks convert to $s$ quarks. The higher densities and temperatures in the core will lead to enhanced neutrino luminosity at late times which may be detectable. Similarly, the decreased pressure support of the core could lead to more compact neutron stars as deduced from the LIGO/VIRGO analysis of gravitational waves from the  GW170817 event \cite{LIGO}.
 However, the two-loop correction introduced in this paper will suppress the $s$-quark content at low chemical potential.  Depending upon chemical potential and temperature, this may inhibit the formation of the CFL phase and  diminish the softening of the EoS relative to an EoS with massless $s$-quarks.  This is consistent with the results of chiral perturbation theory in which the dynamical $s$-quark mass increases, thereby suppressing the formation of the CFL phase  \cite{Buballa16}.
 
 \subsection{Comparison to other works}
 There have been a number of efforts to explore the phase diagram in the literature including that  obtained from Lattice QCD calculations such as \cite{bazavov1, aoki, bazavov2}, chemical freeze-out from relativistic heavy ion collision data \cite{heavyion1, heavyion2, heavyion3, heavyion4} and  in recent NJL models  which explore the impact of the phase transition on the physics neutron stars \cite{njl1, njl2}.  As noted above lattice gauge theory with a finite mass $s$-quark leads to a high temperature phase transition at a temperature in the range of 145 MeV $\le T_c \le$ 170 Mev \cite{Kronfeld12, aoki}, possibly favoring the lower end of the range of $T_c = 154 \pm 9$ as in \cite{bazavov1}.  A study  of the free energy of $2+1$ static quarks in LGT has also indicated that the chiral and deconfinement transitions occur at near the same temperature consistent with the bag-model description here   \cite{bazavov2}.  Studies of the reconstructed phase diagram based upon  the chemical freeze-out from relativistic heavy ion collision data   indicate a critical temperature in the range 150 to180 MeV  \cite{heavyion1, heavyion2, heavyion3, heavyion4}.  This higher temperature is consistent with the effect of adding two-loop corrections and possibly indicates a need for a slightly higher value for the QCD bag constant.  The calculation of the impact of this phase diagram on neutron stars would require a detailed model for the nuclear equation of state at high density.  Nevertheless, the studies  based upon the NLJ nuclear models are quite similar to Figure \ref{fig:6}  \cite{njl1, njl2}.  In Ref.~\cite{njl2} it was proposed that $g$-modes in neutron stars might be a means to detect the QCD transition.  This conclusion does not change in the present work

\section{Conclusion}

We have developed a method for the incorporation of realistic  quark masses into the two-loop contribution to the bag-model thermodynamic potential for quarks with finite mass and  finite chemical potential.  
We show that the method of periodic-regularization for finite quark masses is stable for an interesting window of chemical potential up to near the $s$-quark mass, but diverges as the chemical potential exceeds the quark mass.  Nevertheless, this is an improvement over treatments with massless quarks that can significantly overestimate the pressure contribution from quarks with finite mass at low chemical potential. A rigorous Pad\' e interpolation between the finite mass and massless quark regimes was shown to be unstable to the generation of spurious poles in the interpolation scheme, however, we have shown that lower-order interpolation schemes gives a good representation of the Pad\'e infinite series without the introduction of spurious poles.  Nevertheless, in all the interpolation schemes there is an inherent uncertainty of order 30\%.  Moreover, we have found that it is impractical to implement a consistent polynomial interpolation for the wide range of temperatures and chemical potentials of relevance in practical applications.  We suggest that  a simple logarithmic interpolation between the $s$-quark mass and some multiple (say 2 times the quark mass) is within the uncertainty of the various interpolation schemes and much simpler to implement.
 
  We have shown that  implementing the two loop corrections leads to substantial changes in the QCD phase diagram.   In particular, the two-loop corrections push the phase transition to higher temperatures and densities.  For a fixed bag constant, this will tend to diminish   observable effects from the QCD transition in supernovae and neutron stars.   It will also cause a slight change in the QCD phase transition in the early universe.  This could affect the relic temperatures of  neutrinos (e.g. $\nu_\tau$) that decouple near the temperature of the QCD phase transition.

\begin{acknowledgements}
Work at the University of Notre Dame is supported by the U.S. Department of Energy under Nuclear Theory Grant DE-FG02-95-ER40934. We would like to appreciate both the anonymous referees whose suggestions made this a more motivated article.
\end{acknowledgements}

\end{document}